\documentclass[12pt]{article}
\usepackage{graphicx}
\usepackage{hyperref}
\usepackage{placeins}

\newcommand\pubnumber{SNSN-323-63}
\newcommand\pubdate{\today}

\def\institute{Georg-August-Universit\"at G\"ottingen\\G\"ottingen, Germany\\
}

\usepackage{lineno}

\usepackage{atlasphysics}
\usepackage{atlasparticle}

\def\Title#1{\begin{center} {\Large #1 } \end{center}}
\def\Author#1{\begin{center}{ \sc #1} \end{center}}
\def\Address#1{\begin{center}{ \it #1} \end{center}}

\newcommand\pubblock{\rightline{\begin{tabular}{l} \pubnumber\\
         \pubdate  \end{tabular}}}
\newenvironment{Abstract}{\begin{quotation}  }{\end{quotation}}
\newenvironment{Presented}{\begin{quotation} \begin{center} 
             PRESENTED AT\end{center}\bigskip 
      \begin{center}\begin{large}}{\end{large}\end{center} \end{quotation}}

\def\beq{\begin{equation}}
\def\eeq#1{\label{#1}\end{equation}}
\def\eeqn{\end{equation}}

\def\beqa{\begin{eqnarray}}
\def\eeqa#1{\label{#1}\end{eqnarray}}
\def\eeqan{\end{eqnarray}}

\let\bar=\overbar

\def\Dslash{\not{\hbox{\kern-4pt $D$}}}
\def\dslash{\not{\hbox{\kern-2pt $\del$}}}

\def\msb{{\bar{\ssstyle M \kern -1pt S}}}

\begin{document}
\begin{titlepage}
\pubblock

\vfill
\Title{$\ttbar$ + $Z$ / $W$ / $\ttbar$ at ATLAS}
\vfill
\Author{Clara Nellist, on behalf of the ATLAS Collaboration}
\Address{\institute}
\vfill
\begin{Abstract}
The newest results from the ATLAS Collaboration for the production of a top-quark pair in association with a $Z$ or $W$ boson, and for the production of four top quarks, are summarised in these proceedings. The measurements were performed with 36.1~fb$^{-1}$ of proton-proton collision data from the Large Hadron Collider at a centre-of-mass energy of 13~TeV.\end{Abstract}
\vfill
\begin{Presented}
$11^\mathrm{th}$ International Workshop on Top Quark Physics\\
Bad Neuenahr, Germany, September 16--21, 2018
\end{Presented}
\vfill
\small
\textcopyright 2019 CERN for the benefit of the ATLAS Collaboration.\\
Reproduction of this article or parts of it is allowed as specified in the CC-BY-4.0 license.
\end{titlepage}
\def\thefootnote{\fnsymbol{footnote}}
\setcounter{footnote}{0}

\section{Introduction}

The study of the production of a pair of top quarks in association with a $W$ or $Z$ boson is referred to as $\ttbar{V}$.
The $\ttbar{Z}$ process provides a direct probe of the weak couplings of the top quark.
In the Standard Model of particle physics (SM), the $\ttbar{W}$ process is only possible through initial state radiation from the incoming light quarks, and therefore an enhancement of the $\ttbar{W}$ process could be an indication of Beyond the Standard Model (BSM) physics.
It is also an important background for BSM physics searches with various final states, such as stop searches and electroweak Super Symmetry productions.
Deviations from the SM can be probed with Effective Field Theory.
The $\ttbar{V}$ process is also an important background for $\ttbar{H}$ multi-lepton searches and four-top production.

The measurement of the SM four-top cross section is now becoming possible with current LHC statistics.
Possible enhancements of SM cross sections from new physics through the production of heavy objects in association with a top-quark pair are also possible.
The theoretical SM cross section for four-top production at a centre-of-mass (CoM) of 13~TeV is 9.2~fb at NLO precision~\cite{FourTopsNLO}, which means that $10^{5}$ top-quark pairs are produced at the LHC at $\sqrt{s}=13$~TeV for every four-top event.

The ATLAS experiment~\cite{ATLAS} at the Large Hadron Collider at CERN, is a multi-purpose particle detector with a forward-backward symmetric cylindrical geometry and a near \(4\pi\) coverage in solid angle.

\section{$\ttbar{V}$}

Previous measurements of the $\ttbar{V}$ cross section with the full ATLAS 8 TeV dataset~\cite{ATLAS:ttV8TeV} measured a significance of 5.0$\sigma$ (4.2$\sigma$) over the background-only hypothesis for $\ttbar{W}$ ($\ttbar{Z}$) production.
There were a number of limiting factors for this measurement, including its statistical uncertainty.
For the $\ttbar{W}$ fit, the dominant systematic uncertainty was the modelling of fake leptons and background processes with misidentified charge.
For the $\ttbar{Z}$ fit, the dominant systematic uncertainty source was the modelling of backgrounds from simulation.

Performing the analysis at 13~TeV, instead of 8~TeV, was more advantageous for $\ttbar{Z}$ than for $\ttbar{W}$ due to the rapidly increasing $\ttbar$ background for $\ttbar{W}$ with respect to the more moderate increase in signal cross section.

\subsection{$\ttbar{V}$ at 13 TeV}

\begin{figure}[htb]
\centering
\includegraphics[width=0.48\textwidth]{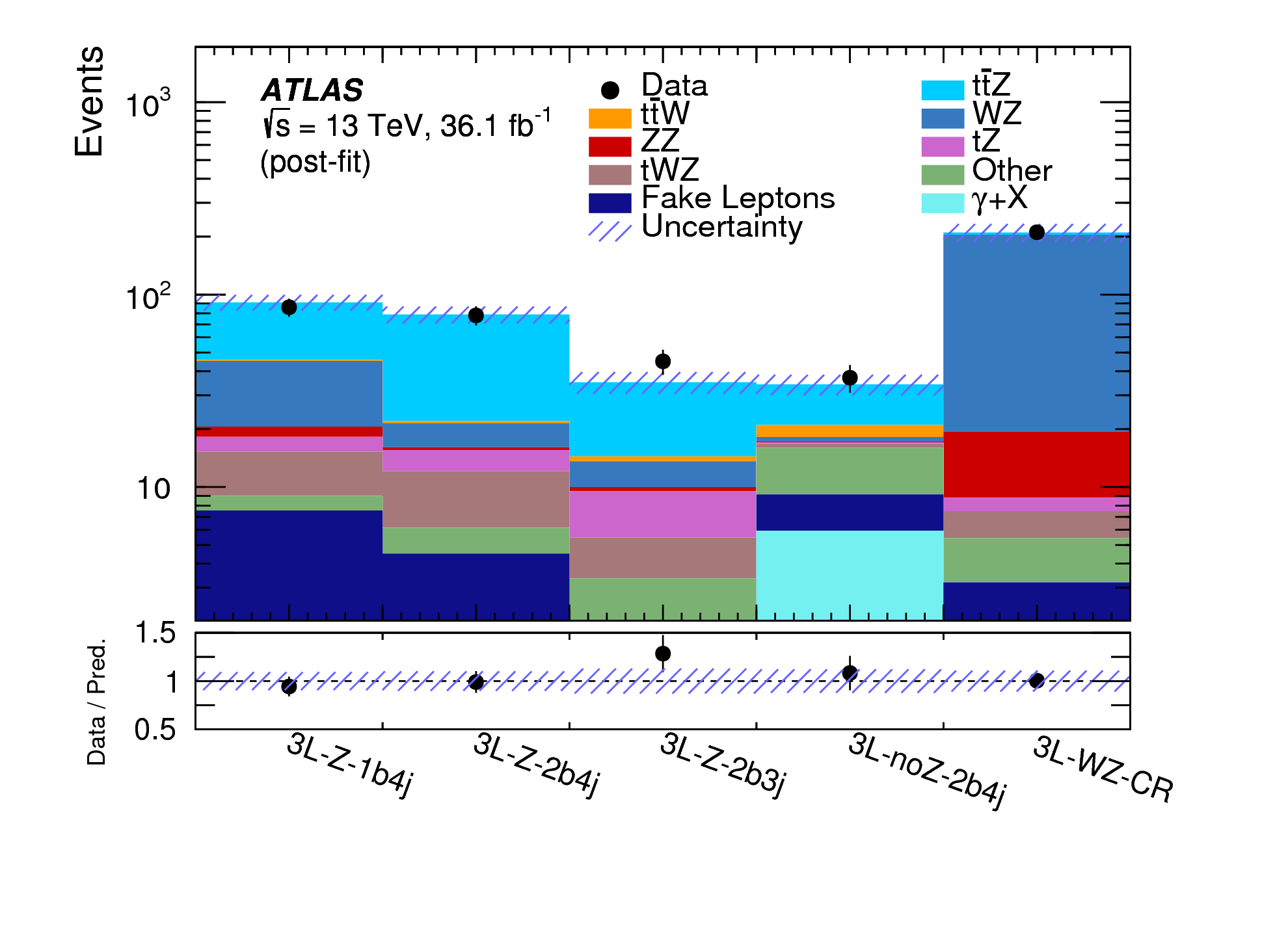}
\includegraphics[width=0.48\textwidth]{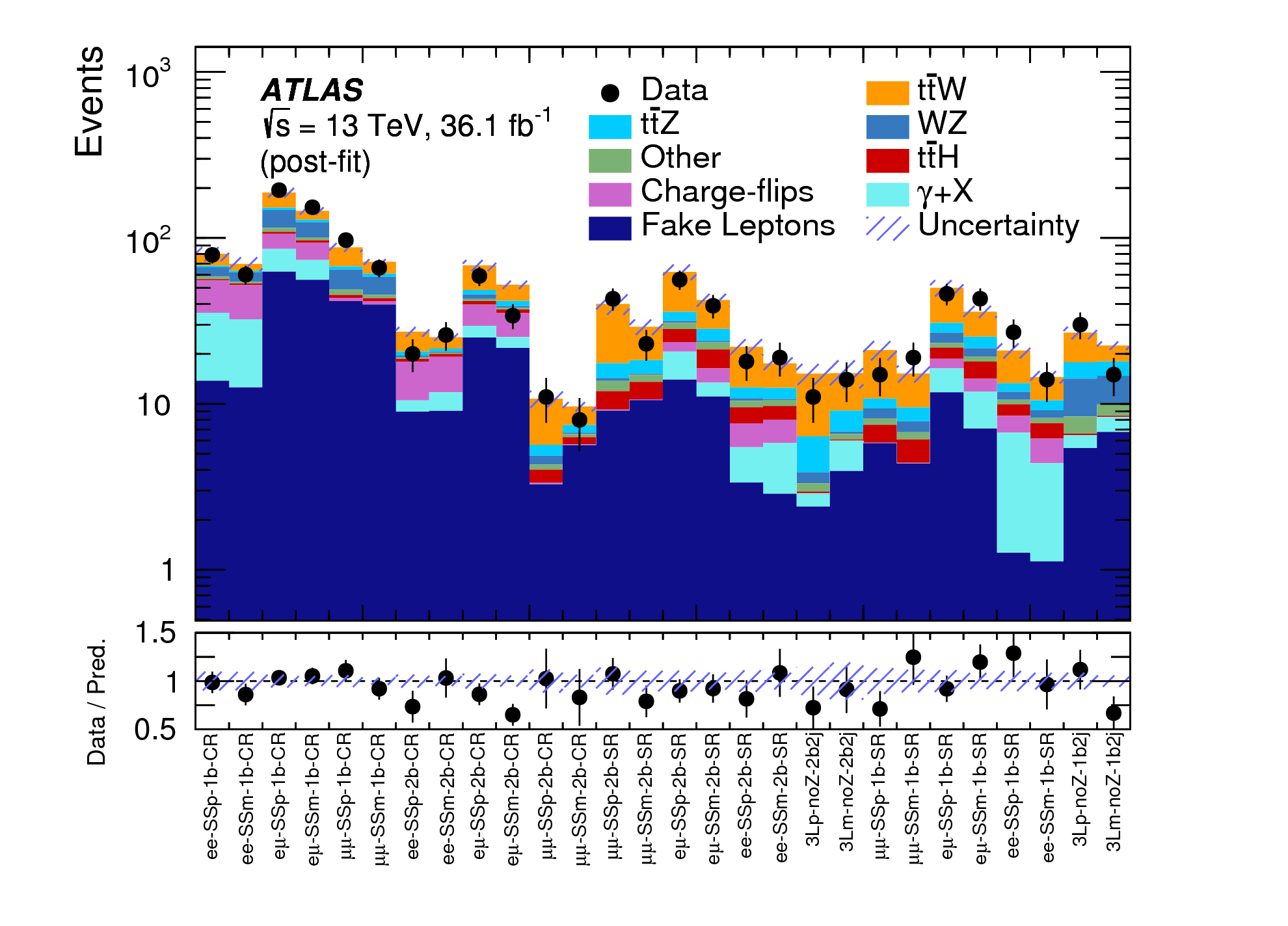}\\
\caption{Event yields in data compared with the fit results in, left, the trilepton signal regions targeting the $\ttbar{Z}$ process and, right, regions targeting the $\ttbar{W}$ process~\cite{ttV13TeV}. }
\label{fig:ttV-sig}
\end{figure}

A simultaneous measurement of the $\ttbar{Z}$ and $\ttbar{W}$ production cross sections in final states with two, three or four isolated electrons or muons, using 36.1 fb$^{-1}$ of data from 2015 and 2016 was performed~\cite{ttV13TeV}.
Each channel is further divided into multiple regions to maximise the sensitivity of the measurement. The signal regions for the channel with three charged leptons in the final state can be seen as an example in Figure~\ref{fig:ttV-sig}~(left), and for regions targeting the $\ttbar{W}$ process in Figure~\ref{fig:ttV-sig}~(right).

Background events containing prompt leptons were estimated from the Monte Carlo (MC) simulation. Normalisation corrections were obtained from the control regions included in the log likelihood fit. 
A data-driven approach was taken for the $\ttbar$ background in the channel with two charged leptons of opposite sign in the final state (2LOS).
The validation region was selected by inverting the requirement that the charged leptons have the same flavour.
Contributions from charge-flip events, significant for events with electrons in the final state, were estimated from data.
This background in events with two muons in the final state was deemed to be negligible since the probability of misidentifying the charge of a muon in the transverse momentum range is very small.
Backgrounds with greater than or equal to one fake charged lepton were modelled using data in dedicated control regions.

\begin{figure}[htb]
\centering
\includegraphics[width=0.43\textwidth]{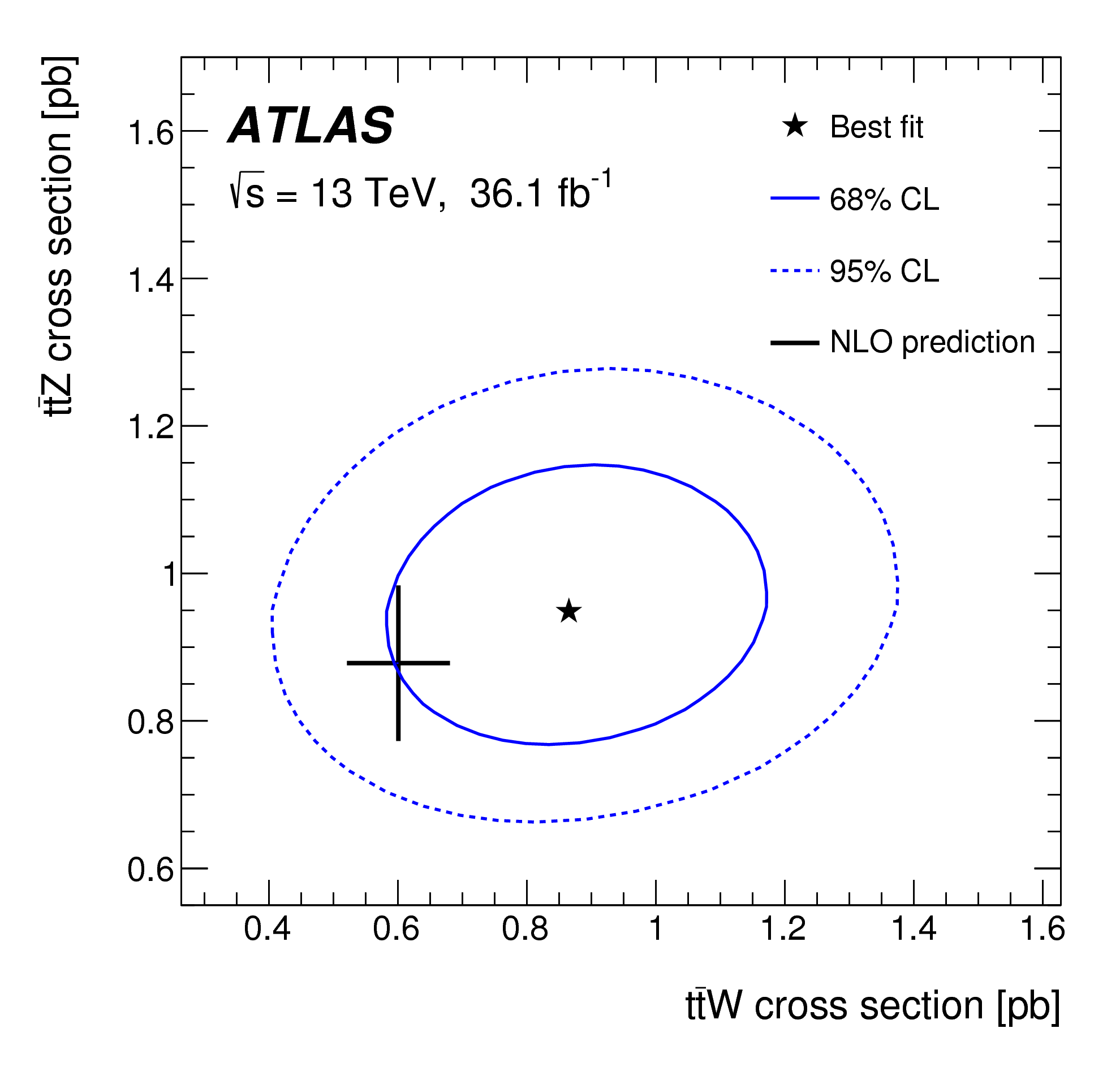}
\caption{The result of the simultaneous fit to $\sigma_{\ttbar{Z}}$ and $\sigma_{\ttbar{W}}$ with the 68\% and 95\% confidence level contours~\cite{ttV13TeV}.}
\label{fig:ttV-fit}
\end{figure}

In the $\ttbar{W}$ analysis and the $\ttbar{Z}$ channel with three charged leptons in the final state (3L), the fake lepton background was estimated using the matrix method.
For the region with four charged leptons in the final state, the semi data-driven fake factor method was used.
Correction factors accounting for potential data / MC simulation discrepancies were extracted in a dedicated control region and are enriched with processes that contain greater than one fake electron or muon.

\subsubsection{$\ttbar{V}$ Simultaneous fit}

The result of the combined fit, shown in Figure~\ref{fig:ttV-fit}, measured 
$\sigma_{\ttbar{Z}}$ = 0.95 $\pm$ 0.13 pb and 
$\sigma_{\ttbar{W}}$ = 0.87 $\pm$ 0.19 pb. 
This yielded a significance of 
$\ttbar{W}$: 4.3$\sigma$ (3.4$\sigma$) observed (expected).
The $\ttbar{Z}$ result has a significance of greater than 5$\sigma$.

A summary of the systematic and statistical uncertainties for the combined fit can be found in Table~\ref{table:ttV}.

\begin{table}[]
\begin{center}
\begin{tabular}{l c c}
\hline
Uncertainty & $\sigma_{{\ttbar}Z}$  & $\sigma_{{\ttbar}W}$   \\
\hline
Total systematic   & 10\% &16\% \\
Statistical   & 8.4\% & 15\% \\
\hline
Total & 13\% & 22\% \\
\hline
\end{tabular}
\caption{Summary of relative systematic and statistical uncertainties in the measured $\sigma_{\ttbar{Z}}$ and $\sigma_{\ttbar{Z}}$ processes from the fit~\cite{ttV13TeV}.}
\label{table:ttV}
\end{center}
\end{table}

\subsubsection{Constraints on BSM}
Interpretations of the inclusive cross section measurement in terms of Effective Field Theory (EFT) were performed.
Constraints were set on the five operators which modify the $\ttbar{Z}$ vertex. The first two enter the $\ttbar{Z}$ vertex as a linear combination, as the measurement is sensitive to the difference. However, only one operator was considered at a time.

\section{Four-tops}

The four top signal is very busy, since there are many jets in the final signature, including those that originated from a $b$-quark.

The previous ATLAS measurement was performed with 3.2~fb$^{-1}$ of data taken at $\sqrt{s}=13$~TeV~\cite{ATLAS:tttt_prev}.
Limits on the observed (expected) upper limit of 21 (16) times the four-top standard model cross section at 95\% confidence level were obtained.
This measurement was performed in the single lepton channel only.

The measurements presented here are with 36.1~fb$^{-1}$ data taken in 2015 and 2016 with $\sqrt{s}=13$~TeV.
The analysis was split into different channels depending on the decay of the $W$~boson (either hadronically or leptonically) from each top quark and combined afterwards.
Broadly the separation is to one charged lepton or two charged leptons of opposite sign (1L \& 2LOS, respectively)~\cite{ATLAS:tttt1L2LOS} and two charged leptons with the same sign or three charged leptons (2LSS \& 3L, respectively)~\cite{ATLAS:tttt2LSS3L}.
Again, charged leptons are only referring to electrons or muons.

\subsection{1L and 2LOS}

Events were categorised according to jet, $b$-tagged jet and mass-tagged reclustered large-$R$ jet multiplicities.
Both channels were dominated by the $\ttbar\bbbar$ background, which is not expected to be modelled well, therefore a data-driven approach was used for its estimation.

For the 1L channel, the background from events with a fake or non-prompt lepton was estimated directly from data using the matrix method.
The method called $\ttbar$~Tag~Rate~Function (ttTRF) was used to estimate $\ttbar$ + jets and gives the probability that a jet is $b$-tagged.

In the 2LOS channel, the background was estimated from the MC simulation.

A simultaneous fit was performed in the twenty signal regions and resulted in an observed (expected) $\sigma$(\ttbar\ttbar) of 47~fb (33~fb) with 95\% CL upper limit.
Consequently, the upper limit on $\sigma_{\ttbar\ttbar}$ was measured to be 5.1 (3.6) times the SM prediction.

For four-top production via an EFT model with a four-top-quark contact interaction, an observed (expected) upper limit on $\sigma$ of 21~fb (22~fb) at 95\% CL was measured.
In this case the SM four-top production was considered as a background.
This result is lower than SM as the contact interaction tends to result in final state objects with larger momenta.

\subsection{2LSS and 3L}

These channels were dominated by lepton fakes and $\ttbar{V}$ background.
Data driven techniques were used for fake and non-prompt lepton backgrounds, which was estimated with the matrix method, and charge misidentification in the 2LSS channel.
Irreducible backgrounds (e.g. $\ttbar{V}$) were modelled by the MC simulation.
This measurement used a cut and count analysis in eight signal regions and six validation regions.
The SM $\sigma_{\ttbar\ttbar}$ upper limit was 69~fb (29~fb) for the observed (expected).

\subsection{Combination}
For the combination, all experimental systematic uncertainties were treated as fully correlated between channels.
Background modelling uncertainties were treated as uncorrelated.
An excess of events over the SM background prediction, excluding the SM $\sigma_{\ttbar\ttbar}$ production, was observed (expected) at 2.8 $\sigma$ (1.0 $\sigma$).
Observed (expected) at a 95\% CL upper limit on $\sigma_{\ttbar\ttbar}$  of 49~fb (19~b).
Corresponding to an upper limit on $\sigma_{\ttbar\ttbar}$ of 5.3 (2.1) times the SM prediction.

\begin{figure}[htb]
\centering
\includegraphics[width=0.45\textwidth]{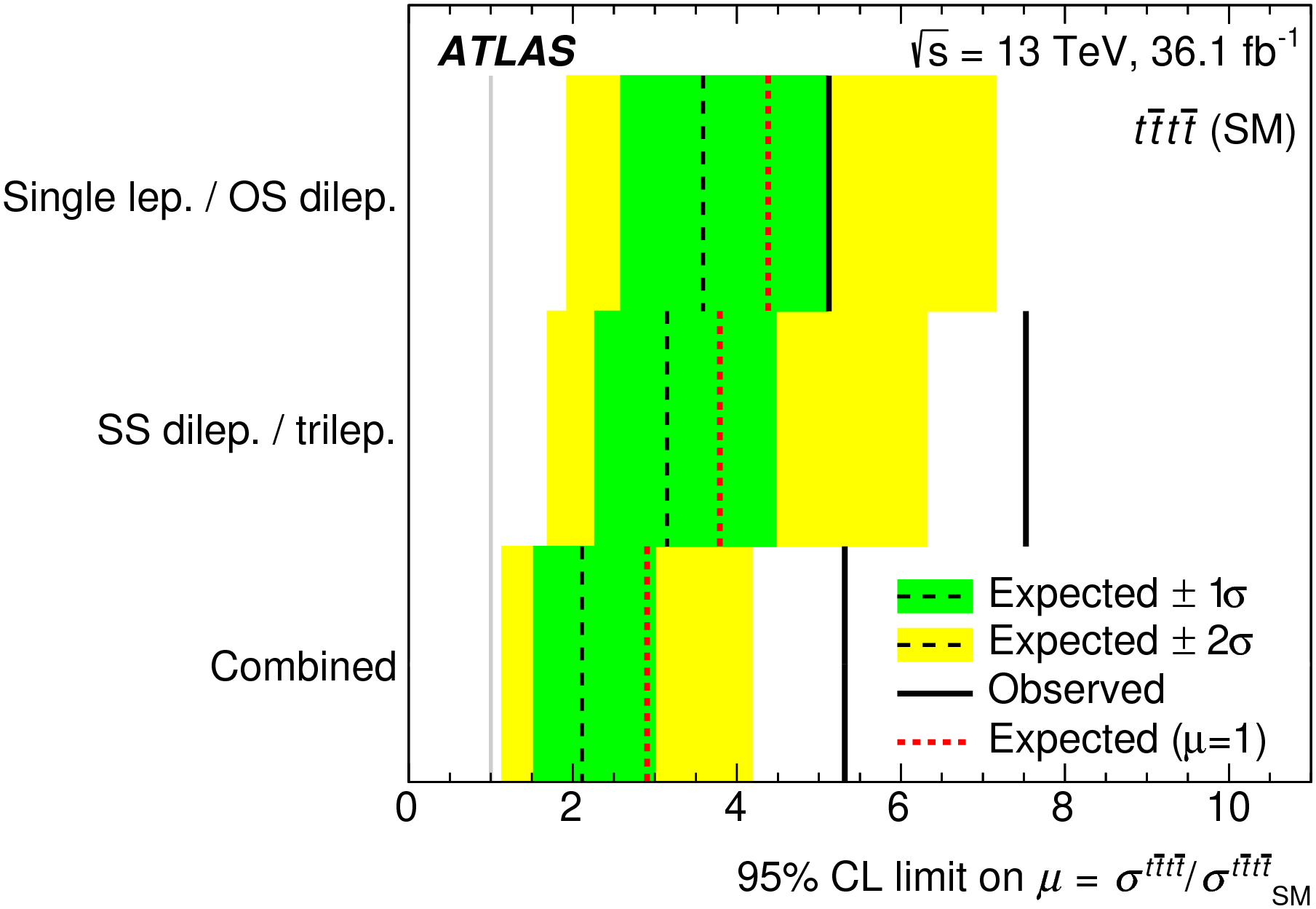}
\includegraphics[width=0.45\textwidth]{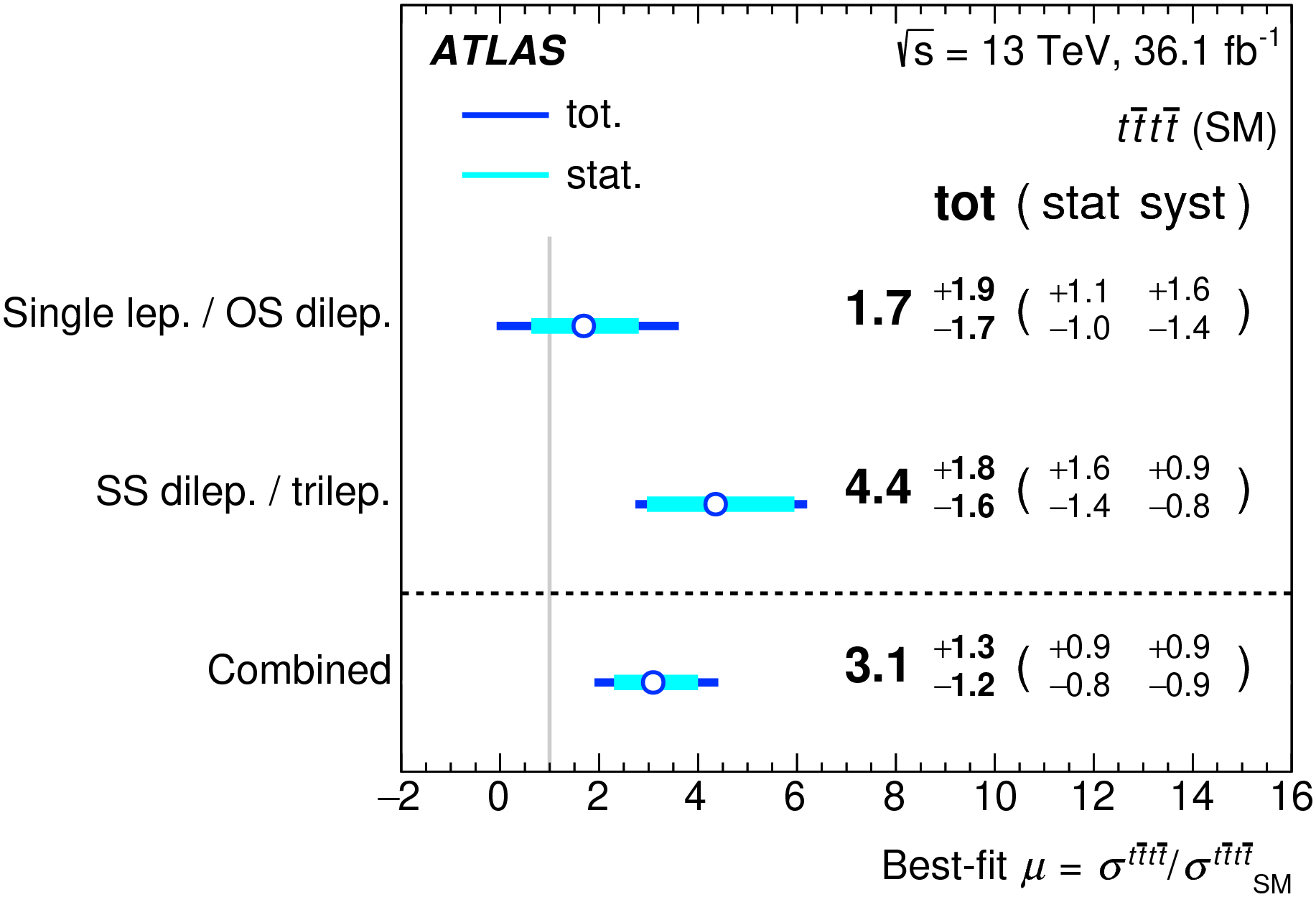}
\caption{Left: The 95\% CL upper limits on $\sigma_{\ttbar\ttbar}$ for the individual measurements and the combined result, relative to the SM prediction~\cite{ATLAS:tttt1L2LOS}. Right: The signal-strength measurements, $\mu$ in the individual channels and for the combination~\cite{ATLAS:tttt1L2LOS}.}
\label{fig:tweets}
\end{figure}

\FloatBarrier

\section{Summary}

Measurement with 36~fb$^{-1}$ 13~TeV data.
Combined fit measured $\sigma_{\ttbar{Z}}$ = 0.95 $\pm$ 0.13~pb
$\sigma_{\ttbar{W}}$ = 0.87 $\pm$  0.19~pb and yields significance of 4.3$\sigma$ (4.3$\sigma$) observed (expected) for $\ttbar{W}$.
The significance of $\ttbar{Z}$ was measured to be greater than 5$\sigma$.
No significant excess above the SM expectation was observed. 
Limits were set on EFT coefficients, which matched or exceeded previous measurements.

The four-tops combined result with all channels using 36~fb$^{-1}$ of 13~TeV data was presented.
An observed (expected) upper limit of 5.3 (2.1) times the four-top SM cross section at 95\% confidence limit was measured.
A slight excess above the SM expectation is observed (expected) at 2.8~$\sigma$ (1.0~$\sigma$), which is driven by the 2LSS / 3L channel.
The kinematic properties of the 2LSS / 3L channel events were compared with expectations from BSM four-tops and found to agree poorly.
The analysis is still statistically dominated; consequently more data is expected to improve the result.




\end{document}